\documentclass[english,pre,a4paper,aps,reprint]{revtex4-2}
\usepackage[T1]{fontenc}
\usepackage[latin9]{inputenc}
\usepackage{amsmath}
\usepackage{amssymb}
\usepackage{stmaryrd}
\usepackage{graphicx}

\makeatletter
\usepackage{natbib}

\makeatother

\usepackage{babel}
\begin{document}
\title{Stochastic entropy production for dynamical systems with restricted
diffusion }
\author{Jonathan Dexter and Ian J. Ford }
\affiliation{Department of Physics and Astronomy, University College London, Gower
Street, London WC1E 6BT, U.K.}
\begin{abstract}
Modelling the evolution of a system using stochastic dynamics typically
implies a greater subjective uncertainty in the adopted system coordinates
as time progresses, and stochastic entropy production has been developed
as a measure of this change. In some situations the evolution of stochastic
entropy production can be described using an Itô process, but mathematical
difficulties can emerge if diffusion in the system phase space is
restricted to a subspace of lower dimension. This can arise if there
are constants of the motion, for example, or more generally when there
are functions of the coordinates that evolve without noise. We discuss
such a case for an open three-level quantum system modelled within
a framework of Markovian quantum state diffusion and show how the
problem of computing the stochastic entropy production in such a situation
can be overcome. We go on to illustrate how a nonequilibrium stationary
state of the three-level system, with a constant mean production rate
of stochastic entropy, can be established under suitable environmental
couplings.
\end{abstract}
\maketitle

\section{Introduction}

Entropy quantifies subjective uncertainty in the configuration of
a system when limited information is available. If the evolution of
such a system is modelled using stochastic dynamics, representing
the effects of coupling to an underspecified environment, then subjective
uncertainty in the configuration of the world (the system plus its
environment) increases with time, corresponding to a growth in the
total entropy. The second law of thermodynamics arises when the world
is characterised or perceived at a coarse grained level, but governed
by underlying equations of motion with a sufficient degree of deterministic
chaos \citep{Ford-book2013}.

For system variables that evolve continuously according to a set of
Markovian stochastic differential equations (SDEs), or Itô processes,
it may be shown that the associated stochastic entropy production
can also be described using an SDE, but only if the noise terms satisfy
certain requirements \citep{SpinneyFord12a,spinney2012entropy,spinney2012use}.
Problems arise from constraints imposed on the dynamics, for example
the existence of constants of the motion. In such cases, the matrix
describing the diffusion of system coordinates in their phase space
becomes singular. There are directions in the space in which diffusion
does not take place and mathematical difficulties in the evaluation
of the stochastic entropy production arise as a consequence.

The central aim of this paper is to show how to take such constraints
on diffusion into account when computing the stochastic entropy production.
In Section \ref{sec:Stochastic-entropy-production} we briefly discuss
how an SDE for stochastic entropy production may be derived from Markovian
SDEs for a set of system coordinates. The mathematical treatment of
cases where diffusion is restricted is discussed in Section \ref{sec:Defining-dynamical-and}.
In Section \ref{sec:An-open-quantum} we consider the stochastic dynamics
of the reduced density matrix of an open three-level quantum system,
subjected to environmental disturbance manifested as raising and lowering
operators associated with transitions between the levels \citep{matos2022,Clarke23}.
The diffusion matrix is singular but we demonstrate how the stochastic
entropy production can still be evaluated. We go on to compute the
environmental component of mean stochastic entropy production numerically
in order to characterise equilibrium and nonequilibrium stationary
states of the system. Application to classical systems is also possible.
Our conclusions are given in Section \ref{sec:Conclusions}.

\section{Stochastic entropy production for Itô processes\label{sec:Stochastic-entropy-production}}

We consider a set of coordinates $\boldsymbol{x}\equiv(x_{1},x_{2},\cdots,x_{N})$
that specify the configuration of a system, and model their evolution
using Markovian stochastic differential equations, or Itô processes:

\begin{equation}
dx_{i}=A_{i}(\boldsymbol{x},t)dt+\sum_{j}B_{ij}(\boldsymbol{x},t)dW_{j},\label{eq:a100-1}
\end{equation}
where the $dW_{j}$ are independent Wiener increments. We define \citep{spinney2012entropy}
\begin{eqnarray}
A_{i}^{{\rm irr}}(\boldsymbol{x},t) & = & \frac{1}{2}\left[A_{i}(\boldsymbol{x},t)+\varepsilon_{i}A_{i}(\boldsymbol{\varepsilon}\boldsymbol{x},t)\right]=\varepsilon_{i}A_{i}^{{\rm irr}}(\boldsymbol{\varepsilon}\boldsymbol{x},t),\qquad\label{eq:a101-1}
\end{eqnarray}
and
\begin{equation}
A_{i}^{{\rm rev}}(\boldsymbol{x},t)=\frac{1}{2}\left[A_{i}(\boldsymbol{x},t)-\varepsilon_{i}A_{i}(\boldsymbol{\varepsilon}\boldsymbol{x},t)\right]=-\varepsilon_{i}A_{i}^{{\rm rev}}(\boldsymbol{\varepsilon}\boldsymbol{x},t),\label{eq:a102-1}
\end{equation}
where $\varepsilon_{i}=1$ for variables $x_{i}$ with even parity
under time reversal symmetry (for example position) and $\varepsilon_{i}=-1$
for variables with odd parity (for example velocity). The notation
$\boldsymbol{\varepsilon}\boldsymbol{x}$ represents $(\varepsilon_{1}x_{1},\varepsilon_{2}x_{2},\cdots)$.
Defining an $N\times N$ diffusion matrix $\boldsymbol{D}(\boldsymbol{x})=\frac{1}{2}\boldsymbol{B}(\boldsymbol{x})\boldsymbol{B}(\boldsymbol{x})^{\mathsf{T}}$,
the Fokker-Planck equation for the probability density function (pdf)
$p(\boldsymbol{x},t)$ is
\begin{equation}
\frac{\partial p}{\partial t}=-\sum_{i}\frac{\partial}{\partial x_{i}}\left(A_{i}p\right)+\frac{\partial}{\partial x_{i}\partial x_{j}}\left(D_{ij}p\right).\label{eq:fpe}
\end{equation}

The stochastic entropy production of the system and its environment,
associated with the stochastic motion described by Eq. (\ref{eq:a100-1}),
is a measure of the difference in probability between pairs of time-reversed
sequences of events, and is defined by \citep{seifertoriginal}
\begin{equation}
\Delta s_{{\rm tot}}=\ln\left(\frac{{\rm Prob(forward\:trajectory)}}{{\rm Prob(backward\,trajectory)}}\right).\label{Delta s}
\end{equation}
This is usually separated into system and environmental contributions:
\begin{equation}
d\Delta s_{{\rm tot}}=d\Delta s_{\text{sys}}+d\Delta s_{\text{env}},\label{eq:sde for Delta s}
\end{equation}
with $d\Delta s_{\text{sys}}=-d\ln p(\boldsymbol{x},t)$. The evolution
of the environmental stochastic entropy production is governed by
\citep{spinney2012use}:\begin{widetext}
\begin{equation}
\begin{split} & d\Delta s_{\text{env}}=-\sum_{i}\frac{\partial A_{i}^{\text{rev}}(\boldsymbol{x})}{\partial x_{i}}dt+\sum_{i,j}\Biggl\{\frac{D_{ij}^{-1}(\boldsymbol{x})}{2}\left(A_{i}^{\text{irr}}(\boldsymbol{x})dx_{j}+A_{j}^{\text{irr}}(\boldsymbol{x})dx_{i}\right)\\
 & -\frac{D_{ij}^{-1}(\boldsymbol{x})}{2}\left(\left(\sum_{n}\frac{\partial D_{jn}(\boldsymbol{x})}{\partial x_{n}}\right)dx_{i}+\left(\sum_{m}\frac{\partial D_{im}(\boldsymbol{x})}{\partial x_{m}}\right)dx_{j}\right)-\frac{D_{ij}^{-1}(\boldsymbol{x})}{2}\left(A_{i}^{\text{rev}}(\boldsymbol{x})A_{j}^{\text{irr}}(\boldsymbol{x})+A_{j}^{\text{rev}}(\boldsymbol{x})A_{i}^{\text{irr}}(\boldsymbol{x})\right)dt\\
 & +\frac{D_{ij}^{-1}(\boldsymbol{x})}{2}\left(A_{j}^{\text{rev}}(\boldsymbol{x})\left(\sum_{m}\frac{\partial D_{im}(\boldsymbol{x})}{\partial x_{m}}\right)+A_{i}^{\text{rev}}(\boldsymbol{x})\left(\sum_{n}\frac{\partial D_{jn}(\boldsymbol{x})}{\partial x_{n}}\right)\right)dt\\
 & +\frac{1}{2}\sum_{k}\Biggl[D_{ik}(\boldsymbol{x})\frac{\partial}{\partial x_{k}}\left(D_{ij}^{-1}(\boldsymbol{x})A_{j}^{\text{irr}}(\boldsymbol{x})\right)+D_{jk}(\boldsymbol{x})\frac{\partial}{\partial x_{k}}\left(D_{ij}^{-1}(\boldsymbol{x})A_{i}^{\text{irr}}(\boldsymbol{x})\right)\\
 & -D_{ik}(\boldsymbol{x})\frac{\partial}{\partial x_{k}}\left(D_{ij}^{-1}(\boldsymbol{x})\left(\sum_{n}\frac{\partial D_{jn}(\boldsymbol{x})}{\partial x_{n}}\right)\right)-D_{jk}(\boldsymbol{x})\frac{\partial}{\partial x_{k}}\left(D_{ij}^{-1}(\boldsymbol{x})\left(\sum_{m}\frac{\partial D_{im}(\boldsymbol{x})}{\partial x_{m}}\right)\right)\Biggr]dt\Biggr\}.
\end{split}
\label{eq: senvbig}
\end{equation}
\end{widetext}It may be shown that the average of $d\Delta s_{\text{sys}}$
over all possible trajectories is related to the incremental change
in Gibbs entropy of the system: $d\langle\Delta s_{{\rm sys}}\rangle=dS_{G}$,
to which boundary terms should be added in certain circumstances \citep{matos2022}.
Computing the environmental stochastic entropy production, on the
other hand, presents particular difficulties if the diffusion matrix
is singular, since the inverse matrix $\boldsymbol{D}^{-1}$ is required
in the above expression. This is the problem we wish to address here..

\section{Defining dynamical and spectator variables \label{sec:Defining-dynamical-and}}

A singular diffusion matrix may be regarded as a consequence of having
fewer independent noise terms than the number of coupled Itô processes.
For example, diffusion might occur on a two dimensional surface within
a three dimensional phase space of system coordinates when the motion
is described by three Itô processes with only two independent Wiener
increments. There is a direction at each point in the phase space
in which there is no diffusive current, which makes the $3\times3$
diffusion matrix singular. These directions lie parallel to spatially
dependent eigenvectors of the diffusion matrix with zero eigenvalues,
to be referred to as null eigenvectors. The obvious solution is to
establish a reduced set of stochastic differential equations that
describe the random evolution of, in this example, two coordinates
on the surface with the third related deterministically to the other
two. We shall denote the stochastically evolving coordinates as `dynamical'
and the remaining coordinates as `spectators'.

It might be possible in simple cases to identify such a reduced set
of coordinates, perhaps by identifying a constant of the motion. However,
as we increase the dimensionality of the phase space and hence the
size of the diffusion matrix, the difficulties in doing so may become
insurmountable. We therefore require a more general treatment of situations
with a singular diffusion matrix. 

Let us consider a system described by $N$ variables $x_{i}$, each
of which evolves stochastically according to

\begin{equation}
dx_{i}=A_{i}dt+\sum_{j=1}^{M}B_{ij}dW_{j},\label{eq: stochdyn}
\end{equation}
where the $dW_{j}$ are $M$ independent Wiener increments. According
to Itô's lemma \citep{gardiner2009stochastic}, the differential of
a function $f$ of these variables can be written

\begin{equation}
df=\sum_{i=1}^{N}\frac{\partial f}{\partial x_{i}}dx_{i}+\sum_{i,j=1}^{N}\frac{\partial^{2}f}{\partial x_{i}\partial x_{j}}D_{ij}dt,\label{eq: itolemma}
\end{equation}
where the elements of the $N\times N$ diffusion matrix are $D_{ij}=\frac{1}{2}\sum_{k=1}^{M}B_{ik}B_{jk}$.
Consider such a function where the stochastic terms in Eq. (\ref{eq: itolemma})
vanish, i.e.

\begin{equation}
\sum_{i=1}^{N}\sum_{j=1}^{M}\frac{\partial f}{\partial x_{i}}B_{ij}dW_{j}=0.
\end{equation}
Taking the square and using $dW_{i}dW_{j}=\delta_{ij}dt$ we find
that

\begin{align}
\sum_{ijkl}\frac{\partial f}{\partial x_{i}}\frac{\partial f}{\partial x_{k}}B_{ij}B_{kl}\delta_{jl}dt & =\sum_{ijk}\frac{\partial f}{\partial x_{i}}\frac{\partial f}{\partial x_{k}}B_{ij}B_{kj}dt\nonumber \\
 & =2\sum_{ik}\frac{\partial f}{\partial x_{i}}\frac{\partial f}{\partial x_{k}}D_{ik}dt=0,
\end{align}
and we therefore have deterministic (noise-free) evolution of $f$
if

\begin{equation}
\left(\boldsymbol{\nabla}f\right)^{\mathsf{T}}\boldsymbol{D}\boldsymbol{\nabla}f=0.
\end{equation}
Such an outcome can arise if $\boldsymbol{\nabla}f$ is an eigenvector
of $\boldsymbol{D}$ with an eigenvalue equal to zero. A matrix is
singular if one or more of its eigenvalues are zero, so we have established
that $\boldsymbol{D}$ is singular if there exists a function $f$
of the stochastic variables $x_{i}$ that evolves deterministically.
For dynamics evolving in an $N$ dimensional phase space under the
influence of $M$ noises where $N>M$, we conjecture that there will
be $L=N-M$ such functions, each associated with one of the null eigenvectors
of $\boldsymbol{D}$. We shall see that this condition allows us to
recast the calculation of the stochastic entropy production to overcome
the problematic singularity of $\boldsymbol{D}$.

\subsection{Identifying constants of motion}

If the deterministic as well as stochastic terms in Eq. (\ref{eq: itolemma})
vanish then $df=0$ and the function $f$ is a constant of the motion:
the evolution of variables, or coordinates, is constrained to a contour
of constant $f$. We can therefore write

\begin{equation}
\boldsymbol{\nabla}f\cdot d\boldsymbol{x}=0,\label{eq: dotproduct}
\end{equation}
meaning that the infinitesimal vector $d\boldsymbol{x}$ specified
by Eq. (\ref{eq: stochdyn}) is tangential to a contour of $f$. Since
$\boldsymbol{\nabla}f$ is also a null eigenvector of the diffusion
matrix, this constraint on $d\boldsymbol{x}$ is conveniently identified
by evaluating the eigenvectors of $\boldsymbol{D}$.

If there are $L=N-M$ constant functions of the coordinates under
the dynamics we can remove $L$ coordinates from the original $N$
dimensional phase space leaving a reduced $M$ dimensional phase space.
Without loss of generality, the first $M$ coordinates in the set
$\left\{ x_{m}\right\} $ with $m=1,\cdots,M$ will be denoted \emph{dynamical
variables} and the remaining $L$ coordinates $\left\{ x_{l}\right\} $
with\emph{ $l=M+1,\cdots,N$ }are designated\emph{ spectator variables}.
The $L$ constants of the motion mean we should be able to write the
spectator variables as functions of the dynamical variables, namely
$\left\{ x_{l}\left(\left\{ x_{m}\right\} \right)\right\} $. The
division into dynamical and spectator variables is arbitrary, but,
as we shall see, some choices are more convenient for computing stochastic
entropy production than others. 

In this labelling scheme, the top left $M\times M$ block of the diffusion
matrix $\boldsymbol{D}$ remains relevant to the stochastic entropy
calculation and will be non-singular, allowing us to use Eq. (\ref{eq: senvbig})
to compute $\Delta s_{{\rm env}}$ associated with the evolution,
but with $i$ and $j$ ranging only between 1 and $M$ rather than
1 and $N$. However, if the elements of this matrix block depend on
spectator variables, we need to take this into account when performing
derivatives with respect to the dynamical variables. To emphasise
this point, we write the components of $\boldsymbol{A},\boldsymbol{D}$
and $\boldsymbol{D}^{-1}$ that appear in Eq. (\ref{eq: senvbig})
to show their dependence on the dynamical and spectator variables
explicitly:

\begin{equation}
\begin{split}A_{i} & \left(\left\{ x_{m}\right\} ,\left\{ x_{l}\left(\left\{ x_{m}\right\} \right)\right\} \right)\\
D_{ij} & \left(\left\{ x_{m}\right\} ,\left\{ x_{l}\left(\left\{ x_{m}\right\} \right)\right\} \right)\\
D_{ij}^{-1} & \left(\left\{ x_{m}\right\} ,\left\{ x_{l}\left(\left\{ x_{m}\right\} \right)\right\} \right),
\end{split}
\end{equation}
and Eq. (\ref{eq: senvbig}) requires us to evaluate derivatives of
$A_{i}$, $D_{ij}$ and $D_{ij}^{-1}A_{j}$ with respect to the dynamical
variables $\left\{ x_{m}\right\} $. 

We consider derivatives of $D_{ij}$ in the following, but the argument
is easily extend to other expressions. We begin by noting that Eq.
(\ref{eq: dotproduct}) can be separated according to dynamical and
spectator variables such that

\begin{equation}
\sum_{m=1}^{M}\alpha_{km}dx_{m}+\sum_{l=M+1}^{N}\alpha_{kl}dx_{l}=0,\label{eq: split}
\end{equation}
where $\alpha_{km}$ and $\alpha_{kl}$ are the dynamical and spectator
components, respectively, of the $k$th null eigenvector of $\boldsymbol{D}$,
with $k=1,\cdots,L$. Therefore

\begin{equation}
\sum_{l=M+1}^{N}\alpha_{kl}dx_{l}=-\sum_{m=1}^{M}\alpha_{km}dx_{m}.\label{spec-dyn}
\end{equation}
Arranging the $\alpha_{km}$ as elements of a rectangular $L\times M$
matrix $Q_{km}$ and the $\alpha_{kl}$ as elements of a square $L\times L$
matrix $P_{kl}$ we have

\begin{equation}
dx_{l}=-\sum_{k,m}P_{lk}^{-1}Q_{km}dx_{m}=\sum_{m=1}^{M}R_{lm}dx_{m},\label{eq: rmatrix}
\end{equation}
where $R_{lm}$ is an element of the $L\times M$ matrix $\boldsymbol{R}=\boldsymbol{P}^{-1}\boldsymbol{Q}$.
Next we write

\begin{equation}
dD_{ij}=\sum_{m=1}^{M}\frac{\partial D_{ij}}{\partial x_{m}}dx_{m}+\sum_{l=M+1}^{N}\frac{\partial D_{ij}}{\partial x_{l}}dx_{l},\label{eq: chainrule}
\end{equation}
and by substituting $dx_{l}$ from Eq. (\ref{eq: rmatrix}) into Eq.
(\ref{eq: chainrule}), we arrive at the following expression for
the derivative of $D_{ij}$ with respect to the dynamical coordinate
$x_{m}$:

\begin{equation}
\frac{dD_{ij}}{dx_{m}}=\frac{\partial D_{ij}}{\partial x_{m}}+\sum_{l=M+1}^{N}\frac{\partial D_{ij}}{\partial x_{l}}R_{lm}.\label{eq: derivatives}
\end{equation}
We were seeking and have identified an additional term on the right
hand side.

To summarise, the described framework allows the computation of entropy
production in cases where the diffusion matrix is singular as a result
of constraints on the dynamics through $L$ constants of motion. The
method employs these constants of motion to reduce the dimensionality
of the phase space across which the system evolves such that the appropriately
reduced diffusion matrix is non-singular. Having carried out this
transformation, Eq. (\ref{eq: senvbig}) can be used to compute entropy
production with derivatives determined according to Eq. (\ref{eq: derivatives}). 

\subsection{Identifying deterministically evolving functions}

We have considered a function $f$ of the stochastic variables $\left\{ x_{i}\right\} $
evolving according to

\begin{equation}
df=\sum_{i}\frac{\partial f}{\partial x_{i}}dx_{i}+\sum_{i,j}\frac{\partial^{2}f}{\partial x_{i}\partial x_{j}}D_{ij}dt,\label{eq: itolemma2}
\end{equation}
and looked at situations where both the deterministic and stochastic
terms in Eq. (\ref{eq: itolemma2}) vanish, making the function $f$
a constant of motion of the dynamics. However, it is only necessary
for the stochastic terms to vanish for there to be a restriction on
the diffusive motion. We now consider the more general case where

\begin{equation}
df=\left(\sum_{i}\frac{\partial f}{\partial x_{i}}A_{i}+\sum_{i,j}\frac{\partial^{2}f}{\partial x_{i}\partial x_{j}}D_{ij}\right)dt\label{eq: detdyn}
\end{equation}
is nonzero. Following earlier arguments, $\boldsymbol{\nabla}f$ is
still a null eigenvector of $\boldsymbol{D}$ and we take each null
eigenvector to correspond to a deterministically evolving function.
To illustrate this, consider a system evolving stochastically in two
dimensions ($x_{1}$,$x_{2}$) with a diffusion matrix that possesses
a single null eigenvector. A function $f(x_{1},x_{2},t)$ evolving
deterministically according to Eq. (\ref{eq: detdyn}) allows us to
reduce the number of variables needed to describe the motion from
two to one. The other becomes a spectator variable. To understand
this better, imagine that at time $t_{0}$ the function $f$ is given
by

\begin{equation}
f(x_{1},x_{2},t_{0})=c_{0},\label{eq: constant}
\end{equation}
where $c_{0}$ is a constant defining a contour of $f$ in the space.
Equation (\ref{eq: constant}) allows us to express spectator variable
$x_{2}$ as a function of dynamical variable $x_{1}$ at time $t_{0}$. 

Between times $t_{0}$ and $t_{1}=t_{0}+dt$, $f$ changes deterministically
by an amount $df_{t_{0}}$ to define a new contour

\begin{equation}
f(x_{1},x_{2},t_{1})=c_{1},
\end{equation}
according to which we can again express $x_{2}$ as a function of
$x_{1}$ at the later time. The evolution of the system is confined
to a sequence of contours of the deterministically evolving function
$f$ such that we can always express $x_{2}$ in terms of $x_{1}.$
We can parametrise the evolution with a single coordinate and thereby
employ a reduced diffusion matrix to compute an entropy production.
The argument easily generalises to an arbitrary number of dimensions. 

\begin{figure}
\begin{centering}
\includegraphics[width=1\columnwidth]{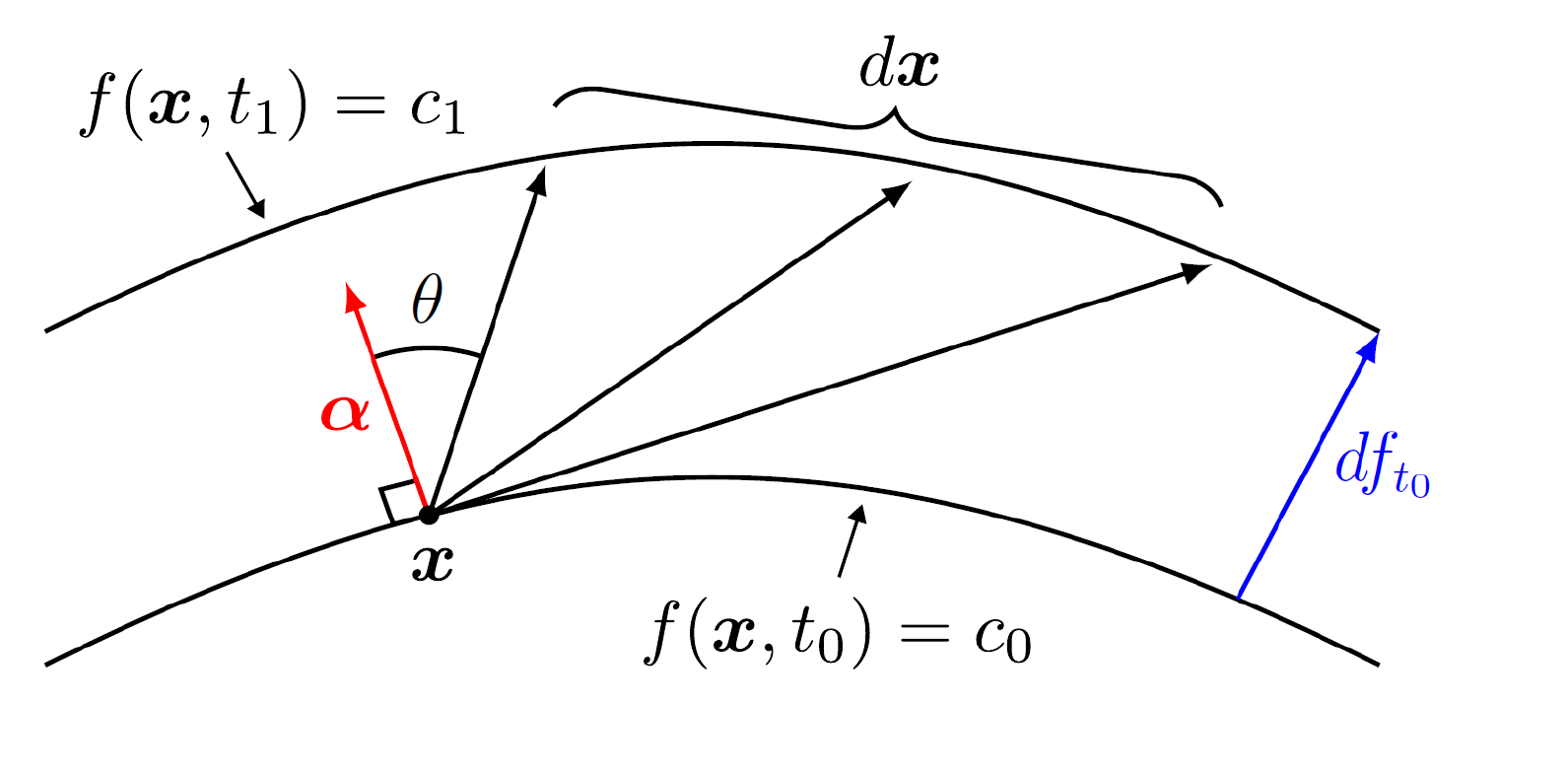}
\par\end{centering}
\caption{Illustration of the evolution of system coordinates between contours
defined by a deterministically evolving function $f(\boldsymbol{x},t)$,
constructed to be normal to a spatially dependent null eigenvector
$\boldsymbol{\alpha}$ of the diffusion matrix. There is a limited
choice of stochastic increments $d\boldsymbol{x}$, defined by angle
$\theta$ and the contours visited at the beginning and end of the
timestep, which restricts the diffusive evolution and complicates
the computation of the stochastic entropy production. \label{fig:contours}}

\end{figure}

As in the previous section we need to consider how derivatives are
modified when we reduce the dimensionality of the phase space. We
write $\boldsymbol{\nabla}f=g\boldsymbol{\alpha}$ where $g$ is equal
to $|\boldsymbol{\nabla}f|$ and $\boldsymbol{\alpha}$ is a normalised
null eigenvector of $\boldsymbol{D}$, , with $|\boldsymbol{\alpha}|=1$,
lying perpendicular to the contour of $f$. Infinitesimal changes
in coordinates $\boldsymbol{x}$ within a timestep $dt$ are given
by $d\boldsymbol{x}$, and $\boldsymbol{x}$ is constrained to pass
between points on specified contours of $f$ at specified times, hence
with only certain $d\boldsymbol{x}$ allowed. The situation is illustrated
in Fig. \ref{fig:contours}.

The component of $d\boldsymbol{x}$ in the direction normal to $f$
is given by

\begin{equation}
d\boldsymbol{x}_{\perp}=|d\boldsymbol{x}|\cos\theta=|\boldsymbol{\alpha}||d\boldsymbol{x}|\cos\theta=\boldsymbol{\alpha}\cdot d\boldsymbol{x},
\end{equation}
where the angle $\theta$ is shown in Fig. \ref{fig:contours}. We
can also write

\begin{equation}
df=|\boldsymbol{\nabla}f|d\boldsymbol{x}_{\perp}=|\boldsymbol{\nabla}f|\boldsymbol{\alpha}\cdot d\boldsymbol{x}=g\boldsymbol{\alpha}\cdot d\boldsymbol{x},\label{eq: df}
\end{equation}
such that 

\begin{equation}
g\boldsymbol{\alpha}\cdot d\boldsymbol{x}=\left(\sum_{i}\frac{\partial f}{\partial x_{i}}A_{i}+\sum_{i,j}\frac{\partial^{2}f}{\partial x_{i}\partial x_{j}}D_{ij}\right)dt.
\end{equation}
For notational convenience we rewrite the term in brackets on the
right hand side as $gG$ so that

\begin{equation}
\boldsymbol{\alpha}\cdot d\boldsymbol{x}=Gdt.
\end{equation}
For a set of null eigenvectors of $\boldsymbol{D}$ labelled by $k=1,...,L$
there are several deterministically evolving functions of the coordinates,
and we can specify relationships between increments in dynamical and
spectator variables such that

\begin{equation}
\sum_{m=1}^{M}\alpha_{km}dx_{m}+\sum_{l=M+1}^{N}\alpha_{kl}dx_{l}=G_{k}dt.
\end{equation}
Following the reasoning in Eq. (\ref{eq: rmatrix}) we write

\begin{equation}
\begin{split}dx_{l} & =-P_{lk}^{-1}Q_{km}dx_{m}+P_{lk}^{-1}G_{k}dt=R_{lm}dx_{m}+S_{l}dt,\end{split}
\label{eq: 1}
\end{equation}
with implied summation over repeated indices, and substituting Eq.
(\ref{eq: 1}) into Eq. (\ref{eq: chainrule}) we obtain

\begin{equation}
dD_{ij}=\sum_{m=1}^{M}\frac{\partial D_{ij}}{\partial x_{m}}dx_{m}+\sum_{l=M+1}^{N}\frac{\partial D_{ij}}{\partial x_{l}}\left(R_{lm}dx_{m}+S_{l}dt\right),
\end{equation}
such that, as before, the required derivatives of the relevant elements
of the diffusion matrix are

\begin{equation}
\frac{dD_{ij}}{dx_{m}}=\frac{\partial D_{ij}}{\partial x_{m}}+\sum_{l}\frac{\partial D_{ij}}{\partial x_{l}}R_{lm}.\label{eq: derivative corrections}
\end{equation}
We therefore find that contributions to derivatives with respect to
dynamical variables, where the expression in question also depends
on spectator variables, are the same whether $f$ is a constant of
the motion or a deterministic function of the dynamics. This is useful
since, in general, we are unlikely to be able to determine if the
singularity of a diffusion matrix is due to the existence of a constant
or a deterministically evolving function.

\subsection{Simple example of restricted diffusive evolution}

To illustrate the above reasoning, consider the two SDEs
\begin{align}
dx_{1} & =\left(1-x_{1}^{2}\right)dW\nonumber \\
dx_{2} & =-\frac{1}{2}x_{2}dt-x_{1}x_{2}dW,\label{2 sdes}
\end{align}
which present a case of restricted diffusive evolution since there
are two Itô processes but only one noise. The $2\times2$ diffusion
matrix for the $(x_{1},x_{2})$ phase space can be shown to be singular.
The reversible deterministic terms $A_{1,2}^{{\rm rev}}$ in both
SDEs are zero. We proceed first by regarding $x_{1}$ as the dynamical
variable and $x_{2}$ as a spectator and use Eq. (\ref{eq: senvbig})
to write 
\begin{equation}
\begin{split} & d\Delta s_{\text{tot}}=-d\ln p_{1}(x_{1},t)+\frac{1}{D_{11}}A_{1}^{\text{irr}}dx_{1}-\frac{1}{D_{11}}\frac{dD_{11}}{dx_{1}}dx_{1}\\
 & +\Biggl[D_{11}\frac{d}{dx_{1}}\left(\frac{1}{D_{11}}A_{1}^{\text{irr}}\right)-D_{11}\frac{d}{dx_{1}}\left(\frac{1}{D_{11}}\frac{dD_{11}}{dx_{1}}\right)\Biggr]dt,
\end{split}
\label{eq: senvbig-x1}
\end{equation}
where $p_{1}(x_{1},t)=\int p(x_{1},x_{2},t)dx_{2}$ and $p(x_{1},x_{2},t)$
satisfies the Fokker-Planck equation associated with Eqs. (\ref{2 sdes}).
Since $A_{1}^{{\rm irr}}=0$ and $D_{11}=\frac{1}{2}(1-x_{1}^{2})^{2}$
we can take derivatives unencumbered by implicit dependence on $x_{1}$
arising from dependence on $x_{2}$. 

However, we could just as well decide to compute the stochastic entropy
production by regarding $x_{2}$ as the dynamical variable and $x_{1}$
as the spectator and write
\begin{equation}
\begin{split} & d\Delta s_{\text{tot}}=-d\ln p_{2}(x_{2},t)+\frac{1}{D_{22}}A_{2}^{\text{irr}}dx_{2}-\frac{1}{D_{22}}\frac{dD_{22}}{dx_{2}}dx_{2}\\
 & +\Biggl[D_{22}\frac{d}{dx_{2}}\left(\frac{1}{D_{22}}A_{2}^{\text{irr}}\right)-D_{22}\frac{d}{dx_{2}}\left(\frac{1}{D_{22}}\frac{dD_{22}}{dx_{2}}\right)\Biggr]dt,
\end{split}
\label{eq: senvbig-x2}
\end{equation}
with $p_{2}(x_{2},t)=\int p(x_{1},x_{2},t)dx_{1}$, $A_{2}^{{\rm irr}}=-\frac{1}{2}x_{2}$
and $D_{22}=\frac{1}{2}x_{1}^{2}x_{2}^{2}$. Since $D_{22}$ depends
on the (current) spectator variable $x_{1}$ we have to employ derivatives
like
\begin{equation}
\frac{dD_{22}}{dx_{2}}=\frac{\partial D_{22}}{\partial x_{2}}+R\frac{\partial D_{22}}{\partial x_{1}},\label{R correction}
\end{equation}
and identify the coefficient $R$ using the (single) null eigenvector
of $\boldsymbol{D}$, which may be shown to be proportional to $(x_{1}x_{2},(1-x_{1}^{2}))^{\mathsf{T}}$.
In this example the dynamics preserve the value of the function $f(t)=x_{1}^{2}(t)+Kx_{2}^{2}(t)-1$
with arbitrary constant $K$, as long as $f(0)=0$ is imposed as an
initial condition (namely the motion is confined to an ellipse). We
can therefore use Eq. (\ref{spec-dyn}) in the form $\alpha_{1}dx_{1}+\alpha_{2}dx_{2}=0$
where $\alpha_{i}$ is the $i$th component of the null eigenvector.
We then obtain
\begin{equation}
x_{1}x_{2}dx_{1}=-\left(1-x_{1}^{2}\right)dx_{2},\label{eq:simple example}
\end{equation}
such that $R=-(1-x_{1}^{2})/(x_{1}x_{2})$. The computation of $\Delta s_{\text{env}}$
using Eqs. (\ref{eq: senvbig-x2}) and \ref{R correction}) can then
proceed. 

The point we are making is that in cases of restricted diffusive evolution,
we can divide the stochastic variables arbitrarily into dynamical
and spectator sets. The implication is that some choices of the division
might be more convenient than others; in the case just considered
it is more sensible to regard $x_{2}$ as a spectator variable rather
than $x_{1}$.

\section{An open three-level quantum system with restricted diffusion\label{sec:An-open-quantum}}

\subsection{SDEs and selection of spectator variables}

\begin{figure}
\begin{centering}
\includegraphics[width=0.9\columnwidth]{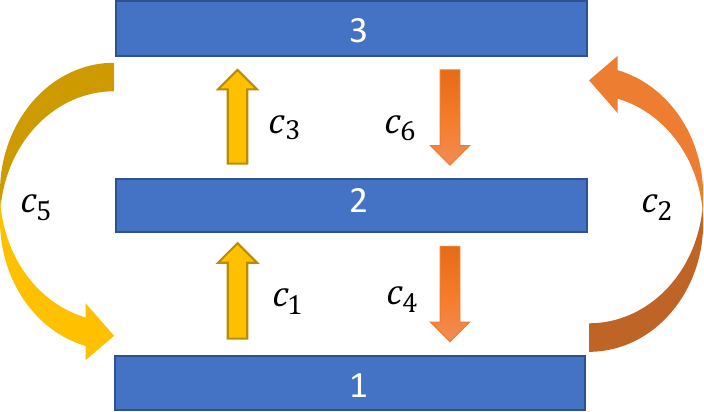}
\par\end{centering}
\caption{A sketch of the dynamics of an open three-level quantum system with
system-environment coupling characterised by the set of six raising
($c_{1-3}$) and lowering ($c_{4-6}$) operators. By reducing the
probabilities of the transitions shown in yellow, through multiplying
the Lindblad operators $c_{1,3,5}$ in the stochastic Lindblad equation
(\ref{ito}) by a weighting factor $w<1$, a probability current through
states $1\to3\to2\to1$ characterised by positive mean stochastic
entropy production can be generated. For $w=1$ the system would otherwise
adopt an equilibrium state with zero current and zero stochastic entropy
production. }
\label{fig: 3 level markov}
\end{figure}

We have developed the present framework for computing stochastic entropy
production because there are physical systems of interest where some
of the stochastically evolving variables are spectators. Specifically,
we consider the dynamics of an open quantum system characterised by
the stochastic evolution of its (reduced) density matrix $\rho$.
The stochasticity is brought about by coupling to the environment,
as described elsewhere \citep{clarke2021irreversibility,matos2022,Clarke23}.
In Appendix \ref{sec:SDEs-for-QSD} it is shown how a Markovian stochastic
Lindblad equation for the evolution of the reduced density matrix
of an open system can be derived starting from the so-called Lindblad
operators that specify the dynamical effect of the environment on
the system. Using this formalism we consider a three-level bosonic
system with environmental coupling characterised by the three raising
($c_{1-3}$) and three lowering ($c_{4-6}$) Lindblad operators given
by
\begin{equation}
\begin{split} & c_{1}=\begin{pmatrix}0 & 0 & 0\\
1 & 0 & 0\\
0 & 0 & 0
\end{pmatrix}\quad\!c_{2}=\begin{pmatrix}0 & 0 & 0\\
0 & 0 & 0\\
1 & 0 & 0
\end{pmatrix}\quad\!c_{3}=\begin{pmatrix}0 & 0 & 0\\
0 & 0 & 0\\
0 & 1 & 0
\end{pmatrix}\\
\\
 & c_{4}=\begin{pmatrix}0 & 1 & 0\\
0 & 0 & 0\\
0 & 0 & 0
\end{pmatrix}\quad\!c_{5}=\begin{pmatrix}0 & 0 & 1\\
0 & 0 & 0\\
0 & 0 & 0
\end{pmatrix}\quad\!c_{6}=\begin{pmatrix}0 & 0 & 0\\
0 & 0 & 1\\
0 & 0 & 0
\end{pmatrix}
\end{split}
\label{eq:lindblads}
\end{equation}
in a basis of kets $\vert1\rangle,\vert2\rangle$ and $\vert3\rangle$
corresponding to the three levels. The SDE describing the dynamics
of the system is given by

\begin{equation}
\begin{split}d\rho & =\sum_{i=1}^{6}\left[\left(c_{i}\rho c_{i}^{\dagger}-\frac{1}{2}\rho c_{i}^{\dagger}c_{i}-\frac{1}{2}c_{i}^{\dagger}c_{i}\rho\right)dt\right.\\
 & \left.+\left(\rho c_{i}^{\dagger}+c_{i}\rho-{\rm Tr}\left[\rho\left(c_{i}+c_{i}^{\dagger}\right)\right]\rho\right)dW_{i}\right].
\end{split}
\label{ito}
\end{equation}
A sketch of the inter-level transitions brought about by the $c_{i}$
is given in Fig. \ref{fig: 3 level markov}.

The reduced density matrix $\rho$ for the open quantum system is
a complex, $3\times3$ Hermitian matrix with a unit trace, corresponding
to eight degrees of freedom. We therefore parametrise $\rho$ in terms
of an eight dimensional vector $\boldsymbol{x}$ evolving as

\begin{equation}
d\boldsymbol{x}=\boldsymbol{A}dt+\boldsymbol{B}d\boldsymbol{W},\label{eq: paramaterisation}
\end{equation}
where $\boldsymbol{A}$ is an eight dimensional vector, $\boldsymbol{B}$
is an $8\times6$ matrix and $d\boldsymbol{W}$ is a six dimensional
vector of independent Wiener increments. It is clear that with fewer
noise terms than SDEs, the diffusive motion will be restricted in
some way. 

\begin{figure}
\includegraphics[width=1\columnwidth]{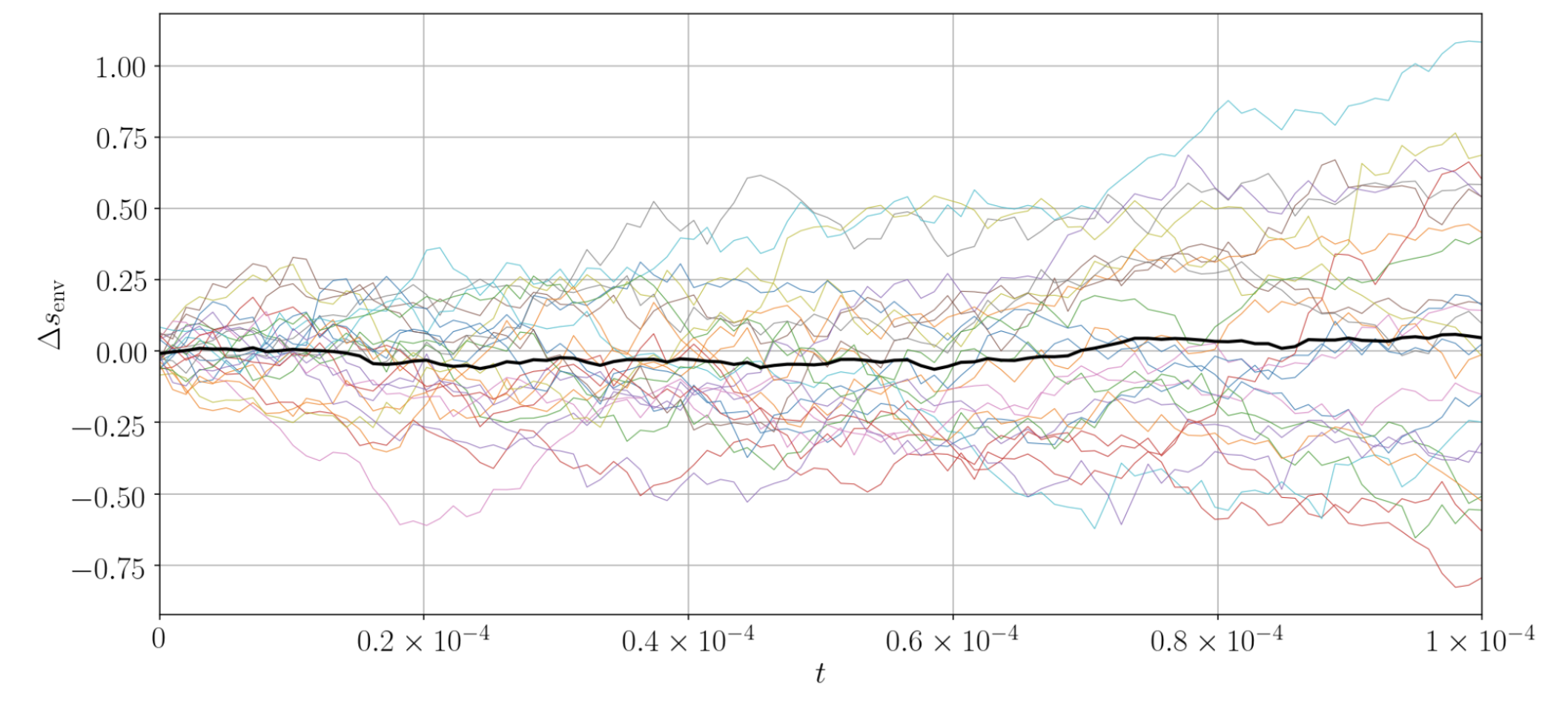} \caption{Environmental component of stochastic entropy production as a function
of time in the three-level quantum system, computed for 25 trajectories
with equally weighted Lindblads. The black line represents the ensemble
mean.}
\label{fig: 3level equal weighting}
\end{figure}

In order to proceed we employ the eight Gell-Mann matrices given by

\begin{equation}
\begin{split} & \lambda_{1}=\begin{pmatrix}0 & 1 & 0\\
1 & 0 & 0\\
0 & 0 & 0
\end{pmatrix}\quad\lambda_{2}=\begin{pmatrix}0 & -i & 0\\
i & 0 & 0\\
0 & 0 & 0
\end{pmatrix}\quad\lambda_{3}=\begin{pmatrix}1 & 0 & 0\\
0 & -1 & 0\\
0 & 0 & 0
\end{pmatrix}\\
\\
 & \lambda_{4}=\begin{pmatrix}0 & 0 & 1\\
0 & 0 & 0\\
1 & 0 & 0
\end{pmatrix}\quad\lambda_{5}=\begin{pmatrix}0 & 0 & -i\\
0 & 0 & 0\\
i & 0 & 0
\end{pmatrix}\quad\lambda_{6}=\begin{pmatrix}0 & 0 & 0\\
0 & 0 & 1\\
0 & 1 & 0
\end{pmatrix}\\
\\
 & \lambda_{7}=\begin{pmatrix}0 & 0 & 0\\
0 & 0 & -i\\
0 & i & 0
\end{pmatrix}\quad\lambda_{8}=\frac{1}{\sqrt{3}}\begin{pmatrix}1 & 0 & 0\\
0 & 1 & 0\\
0 & 0 & -2
\end{pmatrix}.
\end{split}
\label{eq:gell-mann}
\end{equation}
It is convenient to work with eight real variables $s,t,u,v,w,x,y$
and $z$, defined by ${\rm Tr}\left(\lambda_{1}\rho\right)=s$, ${\rm Tr}\left(\lambda_{2}\rho\right)=t$,
${\rm Tr}\left(\lambda_{3}\rho\right)=u$ etc, corresponding to the
eight components of $\boldsymbol{x}$. We can then write $\rho$ as

\[
\rho=\begin{pmatrix}\frac{u}{2}+\frac{\sqrt{3}z}{6}+\frac{1}{3} & \frac{s}{2}-\frac{it}{2} & \frac{v}{2}-\frac{iw}{2}\\
\frac{s}{2}+\frac{it}{2} & -\frac{u}{2}+\frac{\sqrt{3}z}{6}+\frac{1}{3} & \frac{x}{2}-\frac{iy}{2}\\
\frac{v}{2}+\frac{iw}{2} & \frac{x}{2}+\frac{iy}{2} & -\frac{\sqrt{3}z}{3}+\frac{1}{3}
\end{pmatrix}.
\]
\begin{widetext} The SDEs now take the form

\begin{equation}
\begin{pmatrix}ds\\
dt\\
du\\
dv\\
dw\\
dx\\
dy\\
dz
\end{pmatrix}=\begin{pmatrix}-2s\\
-2t\\
-3u\\
-2v\\
-2w\\
-2x\\
-2y\\
-3z
\end{pmatrix}dt+\left(\begin{smallmatrix}-sv+x & -s^{2}-u+\frac{\sqrt{3}z}{3}+\frac{2}{3} & -sx+v & -sv & -s^{2}+u+\frac{\sqrt{3}z}{3}+\frac{2}{3} & -sx\\
-tv-y & -st & -tx+w & -tv & -st & -tx\\
v\left(1-u\right) & s\left(1-u\right) & -x\left(u+1\right) & -uv & -s\left(u+1\right) & -ux\\
-v^{2}-\frac{2\sqrt{3}z}{3}+\frac{2}{3} & -sv+x & -vx & u-v^{2}+\frac{\sqrt{3}z}{3}+\frac{2}{3} & -sv & s-vx\\
-vw & -sw+y & -wx & -vw & -sw & t-wx\\
-vx & -sx & -x^{2}-\frac{2\sqrt{3}z}{3}+\frac{2}{3} & s-vx & -sx+v & -u-x^{2}+\frac{\sqrt{3}z}{3}+\frac{2}{3}\\
-vy & -sy & -xy & -t-vy & -sy+w & -xy\\
\frac{v\left(-3z+\sqrt{3}\right)}{3} & \frac{s\left(-3z+\sqrt{3}\right)}{3} & \frac{x\left(-3z+\sqrt{3}\right)}{3} & -\frac{v\left(3z+2\sqrt{3}\right)}{3} & \frac{s\left(-3z+\sqrt{3}\right)}{3} & -\frac{x\left(3z+2\sqrt{3}\right)}{3}
\end{smallmatrix}\right)\begin{pmatrix}dW_{1}\\
dW_{2}\\
dW_{3}\\
dW_{4}\\
dW_{5}\\
dW_{6}
\end{pmatrix}.\label{big sde}
\end{equation}
The dynamics of $\rho$ are expressed in terms of $N=8$ Itô processes
with $M=6$ noise terms. We therefore expect the diffusion matrix
to be singular (this has been checked using Mathematica) and for there
to exist $L=N-M=2$ deterministically evolving functions of the stochastic
variables $s,\cdots,z$. We choose to assign $y$ and $z$ as spectator
variables allowing us to focus instead on the six SDEs:

\begin{equation}
\begin{pmatrix}ds\\
dt\\
du\\
dv\\
dw\\
dx
\end{pmatrix}=\begin{pmatrix}-2s\\
-2t\\
-3u\\
-2v\\
-2w\\
-2x
\end{pmatrix}dt+\left(\begin{smallmatrix}-sv+x & -s^{2}-u+\frac{\sqrt{3}z}{3}+\frac{2}{3} & -sx+v & -sv & -s^{2}+u+\frac{\sqrt{3}z}{3}+\frac{2}{3} & -sx\\
-tv-y & -st & -tx+w & -tv & -st & -tx\\
v\left(1-u\right) & s\left(1-u\right) & -x\left(u+1\right) & -uv & -s\left(u+1\right) & -ux\\
-v^{2}-\frac{2\sqrt{3}z}{3}+\frac{2}{3} & -sv+x & -vx & u-v^{2}+\frac{\sqrt{3}z}{3}+\frac{2}{3} & -sv & s-vx\\
-vw & -sw+y & -wx & -vw & -sw & t-wx\\
-vx & -sx & -x^{2}-\frac{2\sqrt{3}z}{3}+\frac{2}{3} & s-vx & -sx+v & -u-x^{2}+\frac{\sqrt{3}z}{3}+\frac{2}{3}
\end{smallmatrix}\right)\begin{pmatrix}dW_{1}\\
dW_{2}\\
dW_{3}\\
dW_{4}\\
dW_{5}\\
dW_{6}
\end{pmatrix}.\label{small sde}
\end{equation}
\end{widetext}The diffusion matrix formed from Eq. (\ref{small sde})
via $\boldsymbol{D}=\frac{1}{2}\boldsymbol{B}\boldsymbol{B}^{\mathsf{T}}$
is too elaborate to obtain an analytical expression for its inverse.
It is possible, however, to compute an inverse numerically, remembering
to append terms to any derivatives in the matrix elements according
to the procedure described in the previous section. In actual fact,
with this choice of spectator variables, \emph{no} such additional
terms are required, making this route very convenient.

\subsection{Equilibrium and nonequilibrium stationary states}

Codes have been written to solve the SDEs for the dynamics of the
reduced density matrix and the evolution of the environmental\emph{
}component of the stochastic entropy production. Calculating the system
component $\Delta s_{{\rm sys}}$ requires solution of a Fokker-Planck
equation, which is computationally demanding, but which does not add
to the understanding we develop regarding the \emph{stationary} states
of the system. We comment further on this point later.

The computational demands of solving Eq. (\ref{eq: senvbig}) are
considerable for the system under investigation, so only limited ensembles
of trajectories were generated. Runs for environmental stochastic
entropy production were typically executed with a timestep of $dt=10^{-6}$,
and the reduced density matrix was initiated in the condition $s=t=u=v=w=x=y=z=0.1$
throughout. 

Figure \ref{fig: 3level equal weighting} illustrates the range of
environmental stochastic entropy production for 25 runs. The average
environmental entropy production is nearly always within one standard
deviation of zero, which, given the limited statistics, provides a
strong indication that a zero mean rate of environmental stochastic
entropy production has been established in the stationary state. This
is precisely what is to be expected of an equilibrium state, and the
mean rate of system stochastic entropy production ought to be zero
as well since it corresponds in typical situations to the rate of
change of Gibbs entropy \citep{matos2022}. 

We now investigate a nonequilibrium stationary state with a probability
current passing through the system phase space. Our interest in the
three-level system arises precisely because it is the simplest quantum
system in which such a state might be possible. We create a non-equilibrium
stationary state by breaking detailed balance and favouring a pattern
of transitions $\vert3\rangle\to\vert2\rangle\to\vert1\rangle\to\vert3\rangle$
around the system. We do this by reducing the coupling strength associated
with the Lindblad operators linked to the opposite pattern. Specifically,
we multiply Lindblads $c_{1}$, $c_{2}$ and $c_{5}$ in (\ref{eq:lindblads})
by a weighting factor $w<1$ and derive a modified set of dynamical
equations. Such a non-equilibrium stationary state is expected to
be associated with a positive mean rate of stochastic entropy production.
Moreover, we expect the strength with which the system is weighted
towards such a non-equilibrium stationary state to be related to the
irreversibility of its behaviour and to the degree of mean stochastic
entropy production. 

\begin{figure}
\includegraphics[width=1\columnwidth]{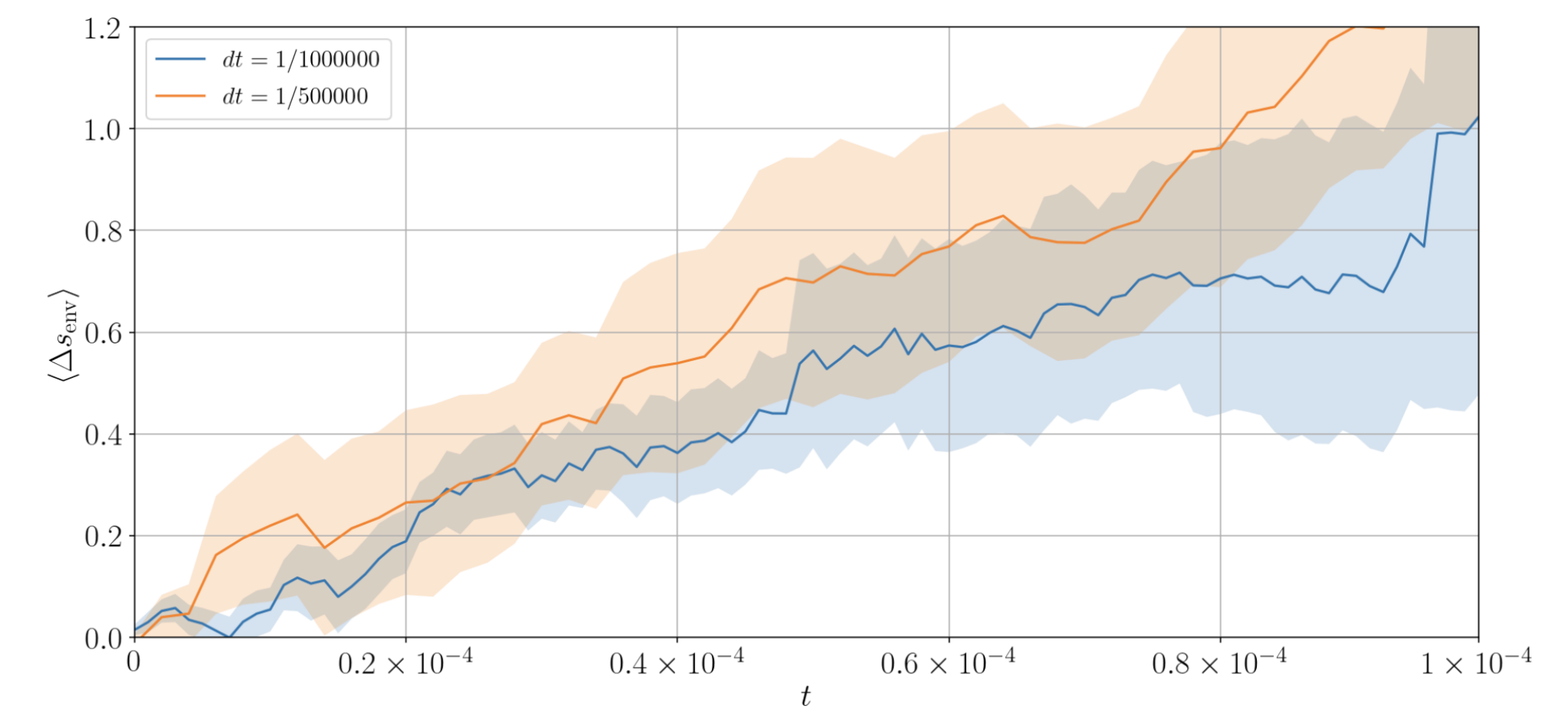} \caption{Comparison of the environmental stochastic entropy production computed
with timesteps $dt=1/1000000$ and $1/500000$ for a weighting of
$w=0.2$. Averages were taken over 50 runs for each value of timestep.
Bands represent the standard error and solid lines the average.}
\label{fig:three level comparison}
\end{figure}

\begin{figure}
\includegraphics[width=1\columnwidth]{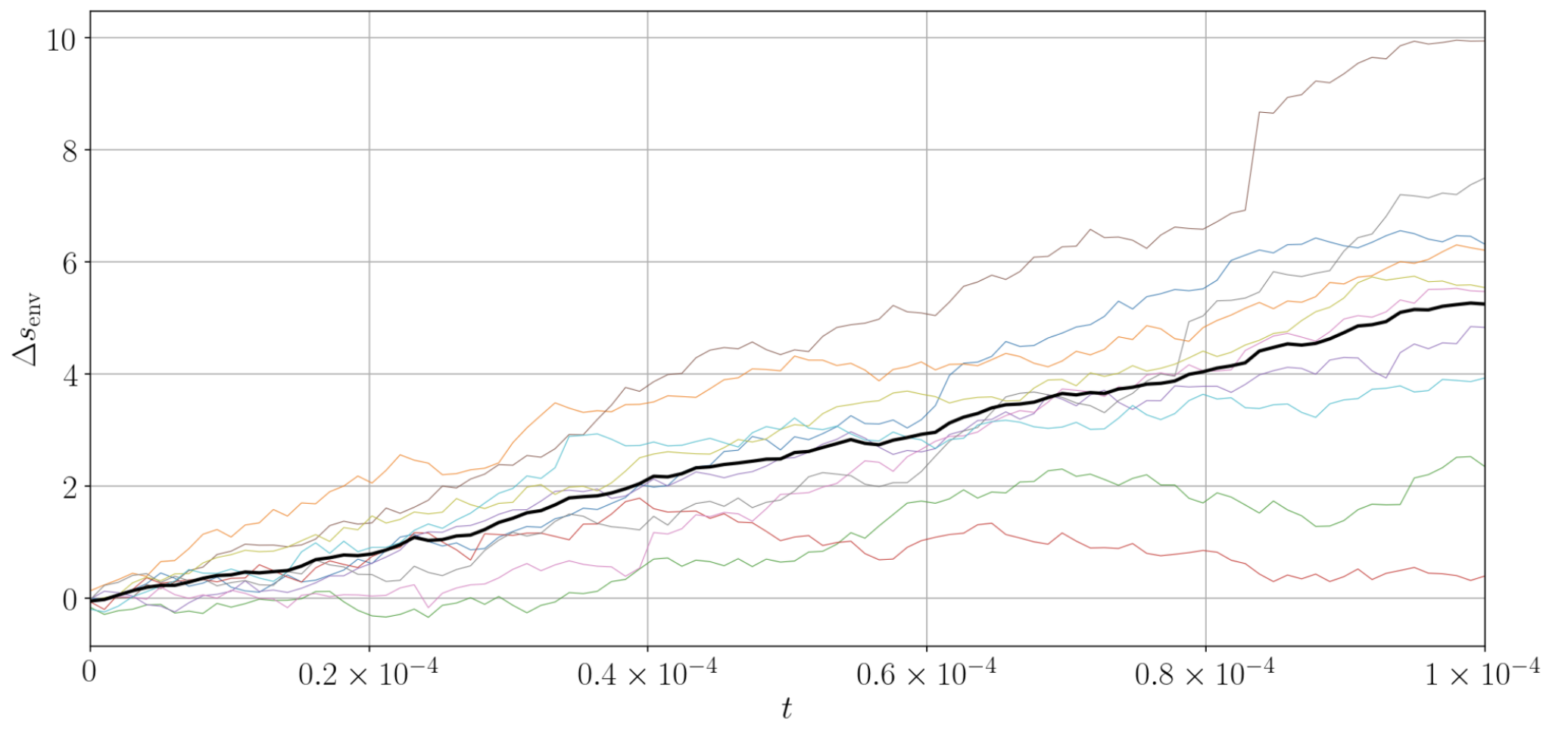} \caption{Environmental stochastic entropy production computed for 10 trajectories
with $w=0.1$ and $dt=10^{-6}$. The black line represents the ensemble
mean.}
\label{fig:0.1 weighting}
\end{figure}

\begin{figure}
\includegraphics[width=1\columnwidth]{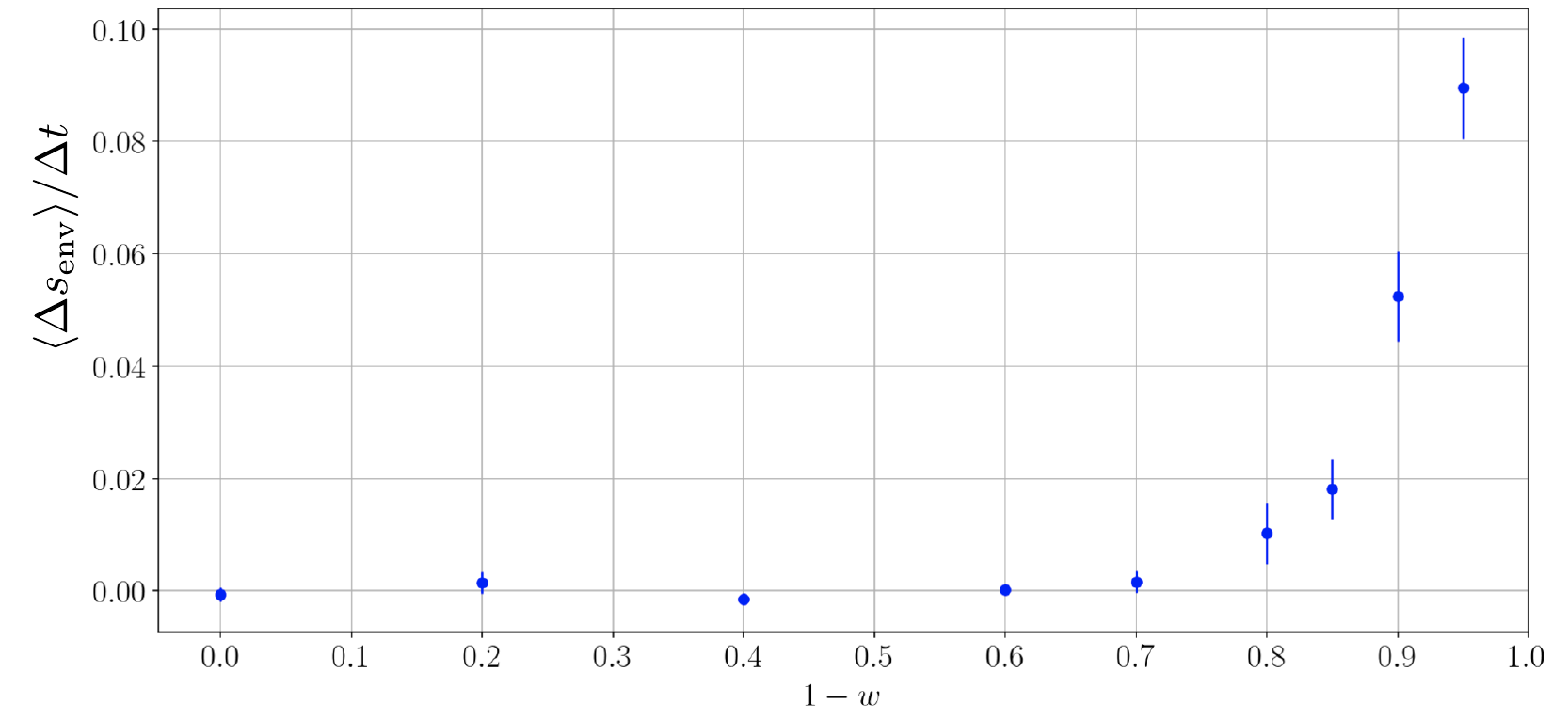} \caption{Mean rate of environmental stochastic entropy production against $1-w$.
The latter characterises the degree to which detailed balance is broken:
a value $w=1$ corresponds to no breakage and the equilibrium state.
Each average was computed by using linear regression to produce best
fit lines for 10 trajectories and then an average was performed over
the gradient of these lines, for a time interval $\Delta t=10^{-4}$.
The error bars show the standard error of these averages.}
\label{fig:weighting comparison}
\end{figure}

We indeed observe a mean positive rate of environmental stochastic
entropy production, within statistical errors. In Fig. \ref{fig:three level comparison}
we check the accuracy of the calculations for $w=0.2$ by comparing
the production for two values of the timestep and find them to be
consistent. Figure \ref{fig:0.1 weighting} shows the environmental
stochastic entropy production for an ensemble of 10 runs with a weighting
of $w=0.1$. 

In line with expectations, we see a constant mean rate of environmental
stochastic entropy production which we associate with the system being
in a nonequilibrium stationary state. The mean system stochastic entropy
production is not expected to make a contribution in a stationary
state since the Gibbs entropy of the system is then constant in time.
Figure \ref{fig:weighting comparison} shows the mean rate of environmental
stochastic entropy production for a selection of weightings. There
is clear indication of a relationship between the breakage of detailed
balance and the mean rate of environmental stochastic entropy production.
For the points $w=0.4$ and $w=0$ we have $\langle\Delta s_{\text{env}}\rangle/\Delta t<0$,
a result at odds with second law, but we attribute this to statistical
error, which could be achieved by, for example, increasing the number
of runs used to generate each point. Figure \ref{fig:weighting comparison}
provides strong support that our approach is a means by which to quantify
the irreversibility of open quantum systems.

\section{Conclusions\label{sec:Conclusions}}

We have employed Itô processes to model the dynamics and thermodynamics
of a system interacting with an environment in the absence of detailed
information about the exact configuration of either. Such an approach
has frequently been used in situations described by classical dynamics
\citep{seifertprinciples}, and recently this has been extended to
quantum systems \citep{matos2022}. In both cases, difficulties arise
when there are fewer independent sources of noise than dimensions
of the system phase space. Diffusion is restricted and the diffusion
matrix becomes singular, which complicates the calculation of stochastic
entropy production.

The solution to the problem is simply to eliminate degrees of freedom
(spectator coordinates) from the entropy calculation to account for
the existence of functions of the coordinates that evolve without
noise. We have described a general method for doing so and illustrated
it for a particle occupying a three-level quantum system, thermalised
by an environment that brings about transitions between the levels.
The dynamics take place in an eight dimensional phase space with only
six noise terms, and the $8\times8$ diffusion matrix is singular. 

We have shown that a stationary equilibrium state of the system may
be established, corresponding to a zero mean rate of environmental
stochastic entropy production (and implicitly by a zero mean rate
of system stochastic entropy production as well). More interestingly,
we have adjusted the probabilities of the transitions induced by the
environment to create a nonequilibrium stationary state as well, where
the particle is made to cycle through the levels in a particular order.
This state is characterised by nonzero mean environmental stochastic
entropy production.

Stochastic entropy production realises Boltzmann's programme of linking
thermodynamics, and specifically entropy production, to the dynamical
evolution of system coordinates \citep{Cercignani98}. Calculating
entropy production quantifies the irreversibility of open system behaviour,
addressing how unlikely it is that reversals of sequences of events
might be observed. Refining this framework to accommodate special
cases such as restricted diffusion adds confidence that such an approach
is the most appropriate tool for quantifying irreversibility in open
classical and quantum systems.

\bibliography{ref}

\appendix

\section{SDEs for quantum state diffusion\label{sec:SDEs-for-QSD}}

The reduced density matrix $\rho$ describing an open quantum system
is considered to evolve stochastically in a timestep $dt$ according
to quantum maps specified by Kraus operators \citep{jacobs2014quantum},
the details of which are provided here. 

We consider $i$ pairs of Kraus operators:
\begin{equation}
M_{i\pm}=\sqrt{\frac{P_{i}}{2}}\left(\mathbb{I}-\frac{1}{2P_{i}}c_{i}^{\dagger}c_{i}dt\pm\frac{1}{\sqrt{P_{i}}}c_{i}\sqrt{dt}\right),\label{eq:a1}
\end{equation}
where $c_{i}$ are the Lindblad operators that specify the mode of
coupling between the system and environment. The meaning of the $P_{i}$
will become clear shortly. A similar scheme has been outlined elsewhere
\citep{clarke2021irreversibility,matos2022,Walls22,Clarke23}, in
which the $P_{i}$ do not appear. 

The Kraus operators satisfy
\begin{align}
M_{i\pm}^{\dagger}M_{i\pm} & =\frac{P_{i}}{2}\left(\mathbb{I}\pm\frac{1}{\sqrt{P_{i}}}(c_{i}+c_{i}^{\dagger})\sqrt{dt}\right),\label{eq:a2-2}
\end{align}
so that $M_{i+}^{\dagger}M_{i+}+M_{i-}^{\dagger}M_{i-}=P_{i}\mathbb{I}$
and the completeness relation \citep{breuer2002theory}
\begin{equation}
\sum_{i}\left(M_{i+}^{\dagger}M_{i+}+M_{i-}^{\dagger}M_{i-}\right)=\mathbb{I},\label{eq:a3}
\end{equation}
holds if $\sum_{i}P_{i}=1$. 

A set of $2i$ reduced density matrices, all of which are positive
definite and of unit trace, are reachable in the timestep starting
from $\rho$, namely
\begin{equation}
\rho^{\prime i\pm}=\frac{M_{i\pm}\rho M_{i\pm}^{\dagger}}{{\rm Tr}\left(M_{i\pm}\rho M_{i\pm}^{\dagger}\right)}.\label{eq:a5}
\end{equation}
The probabilities of making the transitions to these targets are given
by
\begin{align}
p_{i\pm} & ={\rm Tr}\left(M_{i\pm}\rho M_{i\pm}^{\dagger}\right)\nonumber \\
 & =\frac{P_{i}}{2}\left(1\pm\sqrt{\frac{dt}{P_{i}}}{\rm Tr}\left[(c_{i}+c_{i}^{\dagger})\rho\right]\right),\label{eq:a4}
\end{align}
which satisfy $p_{i+}+p_{i-}=P_{i}$ and $p_{i+}-p_{i-}=P_{i}\sqrt{\frac{dt}{P_{i}}}{\rm Tr}\left[(c_{i}+c_{i}^{\dagger})\rho\right]$.
$P_{i}$ is therefore the probability that one of the $i$th pair
of Kraus operators is selected for the map: it is the probability
that the environment disturbs the system in such a way as to transform
$\rho$ into either $\rho^{\prime i+}$ or $\rho^{\prime i-}$. As
a consequence of these dynamical rules, the reduced density matrix
is driven along a Brownian trajectory under the influence of the environment.
Since the Kraus operators reduce to a multiple of the identity as
$dt\to0$, the Brownian path is continuous. No quantum jumps are allowed
(and neither are they necessary \citep{matos2022}).

The statistically averaged form of the new reduced density matrix
is
\begin{align}
\langle\rho^{\prime}\rangle & =\sum_{i}\left(p_{i+}\frac{M_{i+}\rho M_{i+}^{\dagger}}{{\rm Tr}\left(M_{i+}\rho M_{i+}^{\dagger}\right)}+p_{i-}\frac{M_{i-}\rho M_{i-}^{\dagger}}{{\rm Tr}\left(M_{i-}\rho M_{i-}^{\dagger}\right)}\right)\nonumber \\
 & =\sum_{i}\left(M_{i+}\rho M_{i+}^{\dagger}+M_{i-}\rho M_{i-}^{\dagger}\right),\label{eq:a6}
\end{align}
which corresponds in notation with the standard Kraus map \citep{jacobs2014quantum}. 

The average evolution having been established, we now derive an SDE
that describes the stochastic evolution of the reduced density matrix.
The possible increments in the timestep are $d\rho^{i\pm}=\rho^{\prime i\pm}-\rho$
where 
\begin{align}
d\rho^{i\pm} & =\frac{1}{P_{i}}\left(c_{i}\rho c_{i}^{\dagger}-\frac{1}{2}\rho c_{i}^{\dagger}c_{i}-\frac{1}{2}c_{i}^{\dagger}c_{i}\rho\right)dt\nonumber \\
- & \frac{1}{P_{i}}\left(\rho c_{i}^{\dagger}+c_{i}\rho-{\rm Tr}\left[\rho\left(c_{i}+c_{i}^{\dagger}\right)\right]\rho\right){\rm Tr}\left[\rho\left(c_{i}+c_{i}^{\dagger}\right)\right]dt\nonumber \\
\pm & \frac{1}{\sqrt{P_{i}}}\left(\rho c_{i}^{\dagger}+c_{i}\rho-{\rm Tr}\left[\rho\left(c_{i}+c_{i}^{\dagger}\right)\right]\rho\right)\sqrt{dt}.\label{eq:a7}
\end{align}
The average increment in the reduced density matrix is then
\begin{equation}
\langle d\rho\rangle=\sum_{i}\left(p_{i+}d\rho^{i+}+p_{i-}d\rho^{i-}\right),\label{eq:a8}
\end{equation}
which leads after some manipulation to
\begin{equation}
\langle d\rho\rangle=\sum_{i}\left(c_{i}\rho c_{i}^{\dagger}-\frac{1}{2}\rho c_{i}^{\dagger}c_{i}-\frac{1}{2}c_{i}^{\dagger}c_{i}\rho\right)dt.\label{eq:a9}
\end{equation}
The terms in the sum are the dissipators associated with each Lindblad
operator, in the form in which they appear in the standard Lindblad
equation \citep{jacobs2014quantum}.

Next we compute the variance of $d\rho$. For clarity, we write
\begin{equation}
d\rho^{i\pm}=\frac{1}{P_{i}}A_{i}dt\pm\frac{1}{\sqrt{P_{i}}}B_{i}\sqrt{dt},\label{eq:a10}
\end{equation}
where 
\begin{align}
A_{i} & =c_{i}\rho c_{i}^{\dagger}-\frac{1}{2}\rho c_{i}^{\dagger}c_{i}-\frac{1}{2}c_{i}^{\dagger}c_{i}\rho\nonumber \\
 & -\left(\rho c_{i}^{\dagger}+c_{i}\rho-{\rm Tr}\left[\rho\left(c_{i}+c_{i}^{\dagger}\right)\right]\rho\right){\rm Tr}\left[\rho\left(c_{i}+c_{i}^{\dagger}\right)\right],\label{eq:a11}
\end{align}
and $B_{i}=\rho c_{i}^{\dagger}+c_{i}\rho-{\rm Tr}\left[\rho\left(c_{i}+c_{i}^{\dagger}\right)\right]\rho$.
We then construct the average of $\left(d\rho^{i\pm}-\langle d\rho\rangle\right)^{2}$
to lowest order in $dt$. The $A_{i}$ and $\langle d\rho\rangle$
terms do not contribute because they are already of order $dt$ and
we get
\begin{align}
\sum_{i}\left(p_{i+}\left(d\rho^{i+}-\langle d\rho\rangle\right)^{2}+p_{i-}\left(d\rho^{i-}-\langle d\rho\rangle\right)^{2}\right)\nonumber \\
=\sum_{i}\left(p_{i+}\frac{1}{P_{i}}B_{i}^{2}dt+p_{i-}\frac{1}{P_{i}}B_{i}^{2}dt\right)=\sum_{i}B_{i}^{2}dt,\label{eq:a12}
\end{align}
suggesting that the Itô process governing the evolution of $\rho$
is
\begin{align}
d\rho & =\langle d\rho\rangle+\sum_{i}B_{i}dW_{i}\nonumber \\
 & =\sum_{i}\Big[\left(c_{i}\rho c_{i}^{\dagger}-\frac{1}{2}\rho c_{i}^{\dagger}c_{i}-\frac{1}{2}c_{i}^{\dagger}c_{i}\rho\right)dt\nonumber \\
 & +\left(\rho c_{i}^{\dagger}+c_{i}\rho-{\rm Tr}\left[\rho\left(c_{i}+c_{i}^{\dagger}\right)\right]\rho\right)dW_{i}\Big],\label{eq:a13}
\end{align}
and this is the form employed in Eq. (\ref{ito}). 

The Kraus operators in Eq. (\ref{eq:a1}) that underpin our framework
are specified in terms of $P_{i}$ parameters that are to be interpreted
as the probabilities of selection of one of the $i$th pair of Kraus
operators for the transformation of the reduced density matrix. By
design, these parameters do not appear in the SDE, but it is worth
discussing how they might be chosen for use in a Monte Carlo simulation,
for example. If there are $N$ Lindblads, a possible choice could
be $P_{i}=1/N$. This is not completely satisfactory, however, since
it would give equal selection weight to the Kraus operators irrespective
of the degree of coupling between the system and its environment through
each Lindblad operator. It would be better if we ascribe a zero probability
of selection to a Lindblad with zero coupling strength. 

A solution would be to make $P_{i}$ dependent on the norm of the
matrix representing the Lindblad operator. We could use the Frobenius
norm $\bigparallel c_{i}\bigparallel_{\mathrm{F}}=\sqrt{\mathrm{Tr}(c_{i}^{\dagger}c_{i})}$,
for example, and employ probabilities of selection
\begin{equation}
P_{i}=\frac{\bigparallel c_{i}\bigparallel_{\mathrm{F}}}{\sum_{i}\bigparallel c_{i}\bigparallel_{\mathrm{F}}}.\label{eq:a14}
\end{equation}
Adding a null Lindblad $c_{i}=0$ to the set would not change the
Kraus operators representing the other Lindblads, and furthermore
such a null Lindblad would have zero probability of being selected.
The matter is somewhat academic since the $P_{i}$ do not affect the
derived form of the stochastic dynamics, by design, but it does have
impact on the elegance of the framework that emerges for modelling
the quantum state diffusion.
\end{document}